\pdfoutput=1
\documentclass[lettersize,conference]{IEEEtran}
\usepackage{amsmath,amsfonts}
\usepackage{algorithm}
\usepackage{algpseudocode}
\usepackage{array}
\usepackage{textcomp}
\usepackage{stfloats}
\usepackage{url}
\usepackage{verbatim}
\usepackage{graphicx}
\usepackage{cite}
\usepackage{tikz}
\usepackage{subfigure} % Use subfigure with the subfigure option
\usepackage{algpseudocode} 
\usepackage[nolist,nohyperlinks]{acronym}
\usepackage{xcolor}

\usepackage{multirow}
\usepackage{colortbl}
\usepackage{hhline}

\usepackage{tabularray}

\begin{document}

\title{IRQ Coloring and the Subtle Art of Mitigating Interrupt-generated Interference}

%\author{\IEEEauthorblockN{Diogo Costa, Daniel Oliveira, \\José Martins, Sandro Pinto}
%\IEEEauthorblockA{\textit{Centro ALGORITMI / LASI, } \\
%\textit{Universidade do Minho}\\
%Guimarães, Portugal}
%\and
%\IEEEauthorblockN{Luca Cuomo, Ida Maria Savino, \\Bruno Morelli, Fabrizio Tronci, \\Alessandro Biasci}
%\IEEEauthorblockA{\textit{Huawei Pisa Research Center} \\
%Pisa, Italy}
%}

\author{\IEEEauthorblockN{Diogo Costa\IEEEauthorrefmark{1},
Luca Cuomo\IEEEauthorrefmark{2},
Daniel Oliveira\IEEEauthorrefmark{1},
Ida Maria Savino\IEEEauthorrefmark{2},
Bruno Morelli\IEEEauthorrefmark{2},\\
José Martins\IEEEauthorrefmark{1},
Alessandro Biasci\IEEEauthorrefmark{2}, and
Sandro Pinto\IEEEauthorrefmark{1}}
\IEEEauthorblockA{\IEEEauthorrefmark{1}\textit{Centro ALGORITMI / LASI, Universidade do Minho, Portugal}
\IEEEauthorblockA{\IEEEauthorrefmark{2}\textit{Huawei Pisa Research Center, Pisa, Italy}\\
}
}
%\{diogocostaes21, daniel.oliveira, jose.martins\}@dei.uminho.pt}
%\{l.cuomo, i.savino, b.morelli, fabrizio.tronci, a.biasci\}@huawei.com

%\author{Diogo Costa, Luca Cuomo, Daniel Oliveira, Ida Savino, Bruno Morelli, José Martins, Alessandro Biasci, Sandro Pinto~\IEEEmembership{Staff,~IEEE,}
        % <-this % stops a space
\thanks{}% <-this % stops a space
\thanks{Manuscript received April X, X; revised August X, X.}}

\markboth{Journal of \LaTeX\ Class Files,~Vol.~14, No.~8, August~2021}%
{Shell \MakeLowercase{\textit{et al.}}: A Sample Article Using IEEEtran.cls for IEEE Journals}

\IEEEpubid{0000--0000/00\$00.00~\copyright~2021 IEEE}

% Circles with numbers inside
\newcommand*\circled[1]{\tikz[baseline=(char.base)]{
            \node[shape=circle,draw,inner sep=2pt] (char) {#1};}}

\newcommand{\mypara}[1]{\vspace{2pt}\noindent{\textit{\textbf{#1}}}}

% Circles with fill
\setlength{\unitlength}{0.5em}
\newcommand\like[1]{\begin{picture}(1,1)
\ifnum0=#1\put(.5,.35){\circle{1}}\else
\ifnum10=#1\put(.5,.35){\circle*{1}}\else
\put(.5,.35){\circle{1}}\put(.5,.35){\circle*{.#1}}
\fi\fi\end{picture}}

% hatched cell table
\newcommand{\hatch}{\multicolumn{1}{!{\vrule width 1pt\hspace{1pt}}c|}{\cellcolor{white}\slash}}

\maketitle

%%%%%%%%%%%%%%%%%%%%%%%%%%%%%%%%%%%%%%%%%%%%%%%%%%%%%%%%%%%%%%%%%%%%%%%%%%%%%%%%%%%%%%%%%%%%%%%%%%%%
%%%	THIS IS THE ACRONYMS LIST						
%%%%%%%%%%%%%%%%%%%%%%%%%%%%%%%%%%%%%%%%%%%%%%%%%%%%%%%%%%%%%%%%%%%%%%%%%%%%%%%%%%%%%%%%%%%%%%%%%%%%

\begin{acronym}
%#
    \acro{ASIL}{Automotive Safety Integrity Levels}
    \acro{BSP}{Board Support Package}
    \acro{COTS}{Commercial Off-The-Shelf}
    \acro{CPU} {Central Processing Unit}
    \acro{DMA}{Direct memory access}
    \acro{DRAM}{Dynamic Random Access Memory}
    \acro{DTT}{Design-Time Tool}
    \acro{DoS}{Denial-of-Service}
    \acro{FFI}{Freedom From Interference}
    \acro{FPGA}{Field-Programmable Gate Array}
    \acro{GIC}{Generic Interrupt Controller}
    \acro{ICS} {Industrial Control Systems}
    \acro{IRQ}{Interrupt}
    \acro{ISA}{Instruction Set Architecture}
    \acro{LLC}{Last-Level Cache}
    \acro{MCS}{Mixed-criticality Systems}
    \acro{MCU}{Microcontroller Unit}
    \acro{MMU}{Memory Management Unit}
    \acro{OCM}{On-Chip Memory}
    \acro{OS}{Operating System}
    \acro{PID}{Proportional, Integral, Derivative}
    \acro{PMC}{Performance Measuring Counters}
    \acro{PMU}{Performance Monitor Unit}
    \acro{QoS}{Quality of Service}
    \acro{RTM}{Run-Time Mechanism}
    \acro{RTOS}{Real-Time Operating System}
    \acro{SLoC}{Software Lines of Code}
    \acro{SD-VBS}{San Diego Vision Benchmark Suite}
    \acro{SWaP-C}{Size, Weight, Power, and Cost}
    \acro{SoC}{System-on-chip}
    \acro{TCB}{Trusted Computing Base}
    \acro{TLB}{Translation Lookaside Buffer}
    \acro{VM}{Virtual Machine}
    \acro{WCET}{Worst-Case Execution Time}
    
\end{acronym}

% To use them: \ac{Ex}
%%%%%%%%%%%%%%%%%%%%%%%%%%%%%%%%%%%%%%%%%%%%%%%%%%%%%%%%%%%%%%%%%%%%%%%%%%%%%%%%%%%%%%%%%%%%%%%%%%%%

\begin{abstract}
Integrating workloads with differing criticality levels presents a formidable challenge in achieving the stringent spatial and temporal isolation requirements imposed by safety-critical standards such as ISO26262. The shift towards high-performance multicore platforms has been posing increasing issues to the so-called mixed-criticality systems (MCS) due to the reciprocal interference created by consolidated subsystems vying for access to shared (microarchitectural) resources (e.g., caches, bus interconnect, memory controller). The research community has acknowledged all these challenges. Thus, several techniques, such as cache partitioning and memory throttling, have been proposed to mitigate such interference; however, these techniques have some drawbacks and limitations that impact performance, memory footprint, and availability. In this work, we look from a different perspective. Departing from the observation that safety-critical workloads are typically event- and thus interrupt-driven, we mask "colored" interrupts based on the \ac{QoS} assessment, providing fine-grain control to mitigate interference on critical workloads without entirely suspending non-critical workloads. We propose the so-called IRQ coloring technique. We implement and evaluate the IRQ Coloring on a reference high-performance multicore platform, i.e., Xilinx ZCU102. Results demonstrate negligible performance overhead, i.e., $<$ 1\% for a 100 microseconds period, and reasonable throughput guarantees for medium-critical workloads. We argue that the IRQ coloring technique presents predictability and intermediate guarantees advantages compared to state-of-art mechanisms.
\end{abstract}


\section{Introduction}

\ac{MCS} are embedded and/or real-time systems that consolidate workloads with two or more distinct criticality levels (e.g., safety-critical and non-safety-critical) \cite{cerrolaza2020multi, LTZVisor2017, gracioli2019, bao2020}. There are two conflicting requirements in the design of such systems. One relates to the safety guarantees regarding real-time, predictability, and freedom from interference (FFI). The other relates to the need to integrate an ever-growing number of rich functionalities for connectivity and visualization due to the increasing digitalization \cite{cerrolaza2020multi, LTZVisor2017, jailhouse2017}. For instance, it is common to see in modern cars network-connected infotainment systems deployed alongside safety-critical control systems (e.g., anti-lock braking system) \cite{burgio2017software}, while certification requirements (e.g., ISO26262) requiring FFI guarantees in consolidated workloads from different \ac{ASIL} levels.

To cope with this conflicting set of requirements, embedded industries have been resorting to modern high-performance multicore computing platforms endowed with powerful clusters of CPUs, optimized memory hierarchies, and a plethora of application-specific processing units (e.g., GPU, TPU, NPU, FPGA) \cite{gracioli2019, mancuso2013real}. Notwithstanding, it is widely recognized that this exponential complexity and consolidation on multicore platforms have been posing serious challenges for the certification of \ac{MCS}, due to the level of unpredictability and undesired delays \cite{abella2015wcet, cazorla2019probabilistic, cazorla2013proartis} generated by contention at the microarchitectural level, e.g., \ac{LLC}, bus interconnect, and the main memory (DRAM controller).

Interrupts and interrupt-driven workloads further exacerbate this lack of predictability and FFI \cite{martins2023shedding}. Interrupts are (mostly) asynchronous events; thus, they tend to constantly divert the execution flow from the main application logic toward interrupt handlers. Interrupt handlers typically have a completely different code locality, inherently exacerbating the use of shared microarchitectural elements due to the expected \ac{LLC} misses and concurrent accesses to main memory. In a pessimistic scenario, it is reasonable to assume that \ac{DoS} attacks \cite{Bechtel2019} can be constructed with a storm of interrupts triggered, for example, by a bug on a device driver or a malfunction in a particular hardware device.

All these problems are not exotic for the research community. In fact, the real-time system community have acknowledged the issue for quite long time and proposed a set of techniques to minimize such interference. Prominent examples include cache (bank) coloring \cite{kloda2019deterministic, Modica2018, Kim2017}, DRAM bank coloring \cite{Yun2014}, memory throttling \cite{yun2012memory, Modica2018, Farshchi2020}, and I/O regulation \cite{Zini2022, gordon2012eli}. Despite the recognized efforts, existing mechanisms are not perfect in terms of effectiveness and present limitations that impacts performance, memory footprint, and availability. Interestingly, none work focused on interrupts and interrupt-driven workloads as a potential vector steaming interference.

In this work, we propose and reinforce IRQ coloring\footnote{We pioneered the  IRQ coloring concept in our previous workshop paper \cite{Costa2023}; however, (i) the conceptual design was highly simplistic, (ii) the implementation was a minimalist proof-of-concept (tied to the simplistic design), and (iii) the evaluation was very limited and in a synthetic environment} as a novel technique to address interrupt-generated interference and mitigate the effect of cascading failures when FFI cannot be completely guaranteed. The core concept consists of deactivating (or deferring) "colored" interrupts if the \ac{QoS} of critical workloads drops below a specific threshold. By selectively masking interrupts based on the online \ac{QoS} assessment, we provide fine-grained control to mitigate interference on critical workloads without entirely suspending non-critical workloads, i.e., we offer the so-called \textit{intermediate guarantees}. We implemented and evaluated the IRQ Coloring on a real modern Arm high-performance multi-core platform (Xilinx ZCU102) running a static partitioning hypervisor (Bao \cite{bao2020}) and multiple \ac{VM}s. Results for multiple system configurations (i.e., dual and quad-\ac{VM}s) demonstrated negligible overhead (1\%) and reasonable throughput guarantees for medium-critical workloads. We acknowledge predictability and intermediate guarantees advantages compared to state-of-art techniques such as cache coloring and memory throttling.    

In summary, with this paper, we make the following contributions: (i) we provide clear evidence about the impact of interrupt-generated interference supported by empirical experiments with a widely-used benchmark suite (Section \ref{sec:problem-motiv}); (ii) we describe the overall IRQ coloring design, system architecture, and formalization (Section \ref{sec:irq-coloring}); (iii) we discuss the implementation of the technique on a Xilinx ZCU102 platform (Section \ref{sec:irq-coloring}); and (iv) we conduct an extensive evaluation with microbenchmark and benchmark assessment for multiple system configurations (Section \ref{sec:evaluation}). Huawei has already submitted a patent application.

\section{Problem and Motivation} \label{sec:problem-motiv}

\begin{figure*}[t!]
  \centering
  \includegraphics[width=\textwidth]{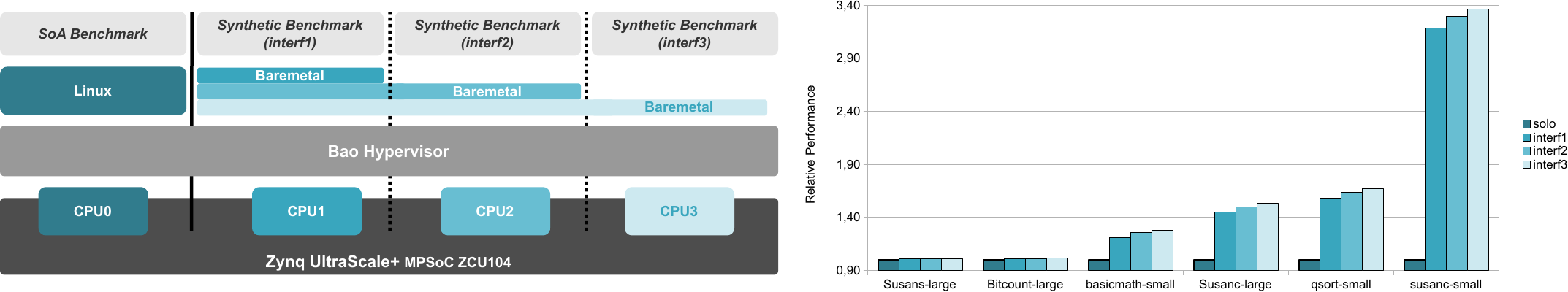}
  \caption{Empirical evidence of interrupt-generated interference. Overview of the setup consisting of one Linux VM running on a single CPU in parallel with a baremetal running on 1, 2, or 3 CPUs (left). Relative performance impact results (right).}
  \label{fig:mibench_profile}
\end{figure*}

Interrupts play a pivotal role in \ac{MCS}, ensuring efficient and timely execution of tasks, especially for safety-critical workloads. When an interrupt occurs (excluding the low-level intricacies of the interrupt entry process), the CPU's execution is redirected to the designated interrupt handler that subsequently runs the event-specific code segment. Interrupts serve the dual purpose of notifying the CPU about incoming events and allowing it to effectively utilize its resources by eliminating the need for continuous polling. Furthermore, interrupts facilitate prompt handling of critical events, such as sensor inputs or actuator commands, ensuring that high-priority tasks are executed within their required deadlines. In safety-critical systems, these benefits, i.e., resource utilization efficiency and real-time responsiveness, are vital for both enhancing the system's performance and preventing catastrophic consequences, as seen in applications such as anti-lock braking systems in automotive designs.

\mypara{Problem: Interrupt-generated interference.} While servicing an interrupt, the CPU deviates from the main execution path to the corresponding interrupt handler. Such diversion typically translates to an entirely different code locality that inherently generates traffic on the microarchitectural shared resources (i.e., excepted \ac{LLC} misses lead to subsequent main memory accesses). Therefore, interrupts can effectively introduce interference on a system and, in some corner cases, lead to a DoS attack due to a storm of interrupts originating from, for example, a buggy device driver or a faulty hardware device. 

\mypara{Evidence: Results from MiBench Automotive Benchmark.} 
To provide sufficient evidence about the interference generated by interrupt handling, we mounted a synthetic use case to conduct some specific experiments. The setup encompasses the Bao hypervisor and two virtual machines (\ac{VM}s): (i) the first \ac{VM} employs a Linux-based system running a selected benchmark retrieved from the MiBench automotive suite \cite{guthaus2001mibench}; (ii) the second \ac{VM} runs a custom baremetal application specifically tailored to pollute the \ac{LLC}. This synthetic application continuously writes to a buffer the same size as the \ac{LLC}. At the same time, its execution is interrupted by a custom hardware module deployed on the FPGA that activates different workloads that write to another segment of the buffer, contributing to creating interference on the shared \ac{LLC}. We have designed four configurations of the experiment, where the first \ac{VM} has assigned one CPU, and the custom baremetal is assigned with one, two, or three CPUs, representing the desired level of interference (\textit{interf1}, \textit{interf2}, and \textit{interf3}, respectively). Figure \ref{fig:mibench_profile} depicts the collected results corresponding to the relative performance of the observed benchmark (i.e., the Linux-based \ac{VM}). The results represent three performance degradation levels resulting from the generated interference. On the bottom side of the spectrum, \textit{bitcount-large} and \textit{susans-large} have a negligible impact (around 1.02x), while \textit{qsort-small} and \textit{susanc-small} have an impact on the execution time of 1.67x to 3.36x, respectively.

\section{IRQ Coloring} \label{sec:irq-coloring}

\begin{figure*}[ht]
  \centering
  \includegraphics[width=0.95\textwidth]{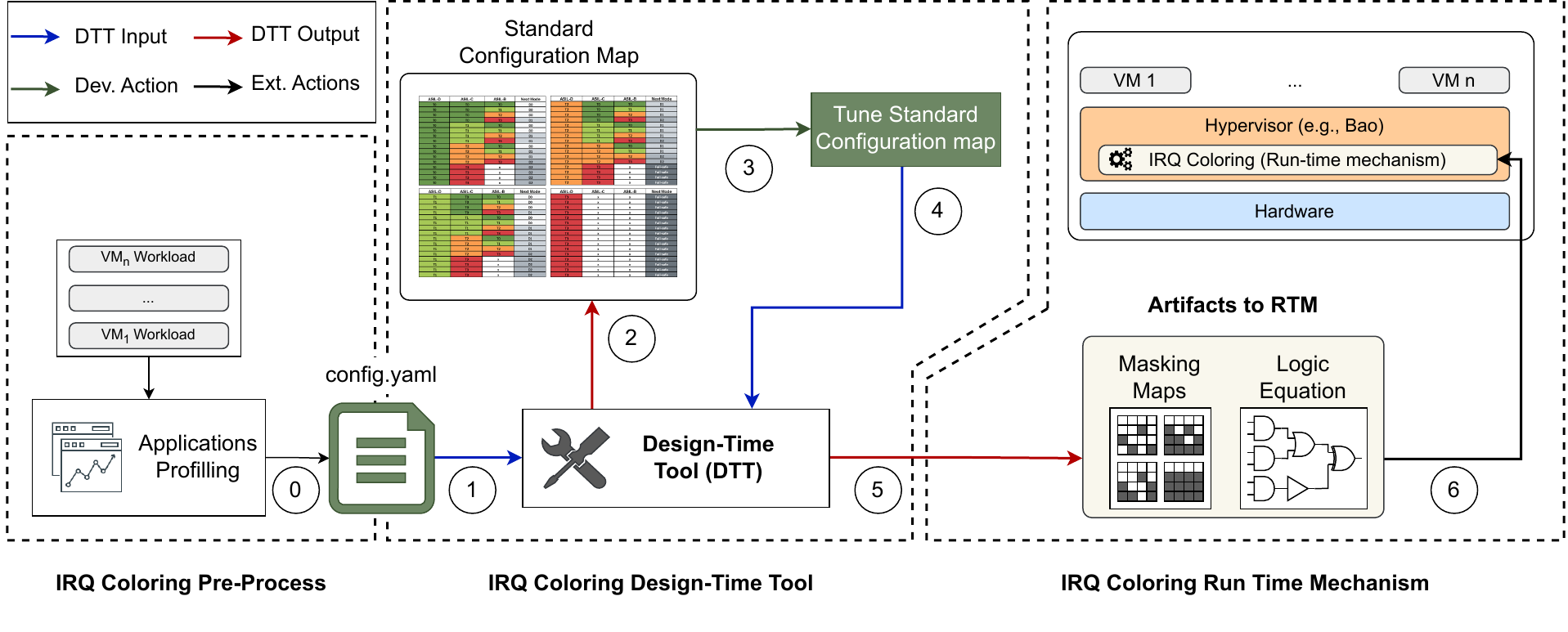}
  \caption{Overview of the IRQ Coloring. On the left, it shows the profiling of workloads of different VMs to specify the configuration file. The middle illustrates the DTT: the input is the configuration file, and the first output is the configuration map; the configuration map can then be tuned to produce the final artifacts for the RTM. On the right, it shows the RTM, which receives the artifacts generated by the DTT and enforces IRQ Coloring at runtime.}
  \label{fig:irqc_sys_overview}
\end{figure*}

The \textit{IRQ Coloring} technique dictates that \ac{IRQ}s assigned to workloads (e.g., \ac{VM}s) are classified according to a specific criticality level and enabled/disabled based on the overall system's performance. The fundamental idea is to selectively deactivate or defer \ac{IRQ}s of non-critical \ac{VM}s if the \ac{QoS} of critical \ac{VM}s drops below a specific threshold. By selectively masking \ac{IRQ}s based on the online assessment of the \ac{QoS} of critical workloads, it is expected that overall interference from non-critical \ac{VM}s to critical \ac{VM}s is mitigated without entirely suspending less critical \ac{VM}s.

The design goals include: (i) guaranteeing near-native \ac{QoS} on higher criticality \ac{VM}s, (ii) maintaining intermediate execution of lower criticality \ac{VM}s (intermediate states under specific degradation modes), and (iii) minimizing the performance impact imposed by the overall mechanism. To achieve these three goals, we have conceived a system with two major artifacts, with the bulk of the logic performed at design time. Figure \ref{fig:irqc_sys_overview} presents the high-level system view, encompassing the \ac{IRQ} Coloring \ac{DTT} and the \ac{IRQ} Coloring \ac{RTM}.

\mypara{0. Pre-processing IRQ Coloring.} The \ac{IRQ} \ac{DTT} requires profiling each interrupt-driven workload to estimate the \ac{WCET} by providing information about execution time and microarchitectural events, such as caches and bus accesses \circled{0}. Different techniques can be employed to profile \ac{VM}s' workload, including (i) profiling, (ii) static analysis, (iii) timing analysis, and (iv) model-based analysis. Therefore, the \ac{IRQ} Coloring \ac{DTT} is agnostic to the profiling technique, allowing developers to choose a different solution.

\mypara{1. IRQ Coloring Design-Time Tool (IRQ DTT).} Based on the established workload profile and assigned \ac{IRQ}s, along with the specification of \ac{VM} criticality \circled{1}, the \ac{IRQ} Coloring \ac{DTT} produces a configuration table (representing the masking map to be applied in each degradation mode) \circled{2} that can be optimized by the user to meet different safety specifications \circled{3}. The \ac{DTT} then leverages the configuration map \circled{4} to create artifacts that feed the \ac{IRQ} \ac{RTM} \circled{5}, at the hypervisor level \circled{6}. These artifacts include masking maps for each degradation mode and a logical equation that defines which degradation mode should be applied.

\mypara{2. IRQ Coloring Run-Time Mechanism (IRQ RTM).} The \ac{IRQ} \ac{RTM} mainly collects specific metrics from the hardware performance counters, i.e., \ac{PMU}, and based on the artifacts produced by the \ac{IRQ} \ac{DTT}, selectively disables \ac{IRQ}s based on the performance of each \ac{VM}. The mechanism consists of three stages: (i) calculating the \ac{QoS} of each \ac{VM}; (ii) decoding the calculated QoS (i.e., creating a 2-bit representation of the \ac{QoS} value); and (iii) applying the control logic equation. This sequential process selects the appropriate degradation mode to be applied to the system, which leverages the masking maps generated by the \ac{DTT} to disable different sets of interrupts and reduce interference.

\subsection{System Architecture}    %%%%%%%%%%%%%%%%%%%%%%%%%%%%%%%%%%%%%%%%%%%%%%%%%%%%%%%%%%%%%%%%%%%%%%%%%%%%%%%%%%%%%%%%%%%%%%%%%%%%%%%%%%%%%%%%%%%%%%%%%%%%%%%%%%%%%%%%%%%%%%%%%%%%%%%%%%%%%%%%%%%%%%%%%

Our solution for mitigating interference in a system with coexisting \ac{VM}s involves modifying the interrupt controller routing system (hypervisor level), leveraging the \ac{IRQ} Coloring \ac{RTM} to mask interrupts for \ac{VM}s. The \ac{GIC} manages interrupt routing, consisting of two primary components: the distributor and the \ac{CPU} interfaces. The distributor routes injected interrupts, while the \ac{CPU} interfaces connect each core to the distributor. The \ac{IRQ} Coloring \ac{RTM} operates between both components, as shown in Figure \ref{fig:irqc_sys_arch}, and is responsible for masking specific \ac{IRQ}s. By masking interrupts assigned to different \ac{VM}s, it is possible to mitigate interference generated by coexisting \ac{VM}s with varying levels of criticality.

In this paper, we specifically focus on the interference generated by interrupt-driven workloads rather than the inner interference resulting from the contention at the GIC level. We also assume that the workload running on each VM consists of a set of applications controlled by different interrupts. Therefore, masking an interrupt would result in suspending a particular application.

\begin{figure}[t!]
  \centering
  \includegraphics[width=\columnwidth]{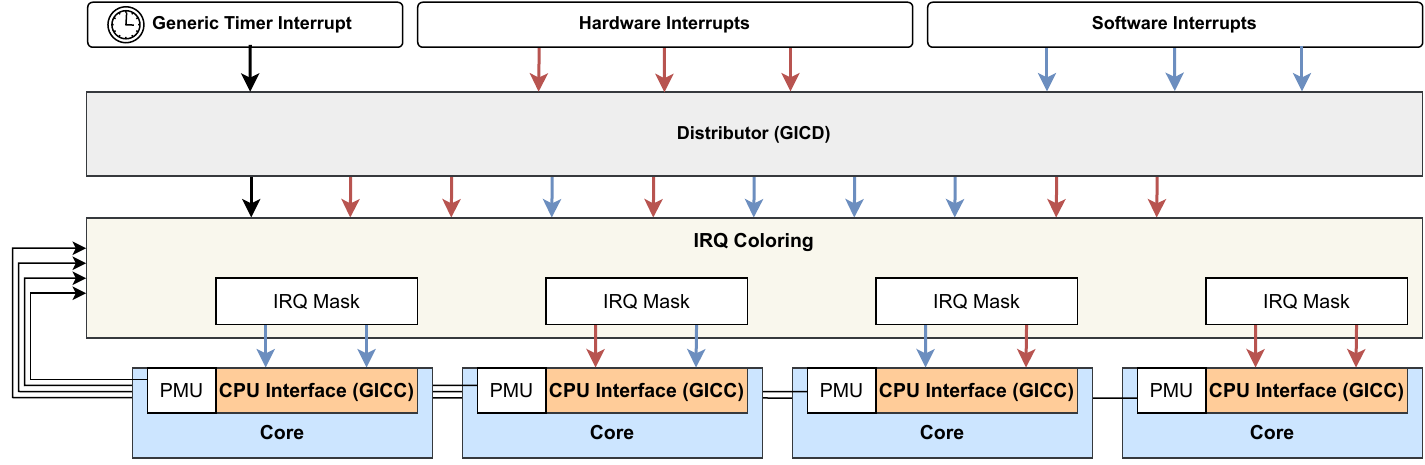}
  \caption{IRQ Coloring RTM system architecture}
  \label{fig:irqc_sys_arch}
\end{figure}

\subsection{IRQ Coloring Formalization} %%%%%%%%%%%%%%%%%%%%%%%%%%%%%%%%%%%%%%%%%%%%%%%%%%%%%%%%%%%%%%%%%%%%%%%%%%%%%%%%%%%%%%%%%%%%%%%%%%%%%%%%%%%%%%%%%%%%%%%%%%%%%%%%%%%%%%%%%%%%%%%%%%%%%%%%%%%%%%%%%%%

The \ac{GIC} is responsible for managing and forwarding the IRQs generated (either by software or hardware) to the respective \ac{VM}s\footnote{Although the overall IRQ Coloring technique description and formalization assume underlying virtualization support, all the concepts are generalizable for other system configurations, mainly when there is a single \ac{OS} instance.} to which they are assigned. In this sense, each IRQ propagated to a given VM ($VM_i$) is identified by a unique index ($k$) - depicted in Figure \ref{fig:irqc_sys_arch} - with the following rules:
\begin{subequations}
\label{eq:form1}
    \begin{equation}
        0 < i < M
    \end{equation}
    \begin{equation}
        0 < k < N
    \end{equation}
\end{subequations}

where $M$ is the max number of \ac{VM} and $N$ is the max number of interrups assigned to a given \ac{VM}.

Each \ac{VM} is associated with a vector ($\hat{P}$) that contains a set of parameters (e.g., number of cache accesses) that allows to calculate the effect of the interference of a given set of events on a given \ac{VM} ($e_{VM_i}$). Such effect is calculated by combining the interferences that affect the target \ac{VM} and each interference is due to the corresponding interference parameter as outlined in Equation \ref{eq:form2}.

\begin{equation}
    \begin{split}
        e_{VM_i} = & f_i(I_{0_{VM_i}}(\hat{P}_{0_{VM_i}}), \\
         I_{1_{VM_i}}(\hat{P}_{1_{VM_i}}), &\ ..., I_{N_{VM_i}}(\hat{P}_{N_{VM_i}}))
    \end{split}
    \label{eq:form2}
\end{equation}

For each interrupt routed to a given \ac{VM}, an IRQ index (IRQk) is defined, which corresponds to the tuple composed of the physical $PIN$ associated with the interrupt and the identifier of the \ac{VM} that manages such interrupt, in which:
\begin{itemize}
    \item $\hat{P}_{j_{VM_i}}$ is the value of the $j$-th parameter that influences the $i$-th VM;
    \item $I_{j_{VM}}(.)$ is the function that calculates the interference caused by the $j$-th parameter on the $i$-th VM;
    \item $f_i(.)$ is the function that combines the different type of interferences that affect the i-th VM.
\end{itemize}

At run-time, the hypervisor monitors the performance of the different \ac{VM}s. If the hypervisor detects any degradation in the higher criticality \ac{VM}s, it switches to a degraded state. To mitigate the interference generated by interrupt-driven workloads the hypervisor masks the IRQs that have the higher impact on the degradation effect ($e_{VM_i}$). If the monitored parameters of the target \ac{VM} continue to indicate non-negligible interference, the hypervisor progressively masks the IRQs associated with a lower degradation effect.
When the hypervisor restores the status of the target \ac{VM}, it progressively unmasks the IRQs starting from the IRQs associated with the lowest degradation effect. In other words, the IRQs are restored in reverse order from the previous points. The hypervisor follows this approach until all IRQs have been unmasked and the target \ac{VM} has been fully restored to its previous state.

\begin{figure}[t]
  \centering
  \includegraphics[width=\columnwidth]{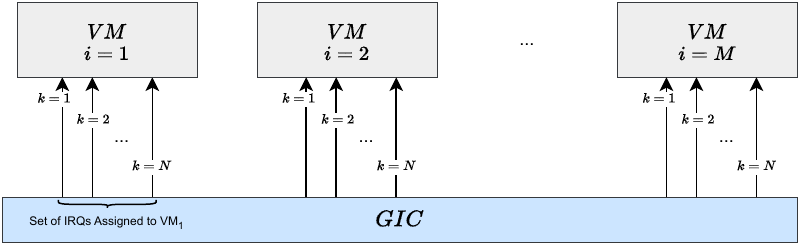}
  \caption{IRQ Coloring RTM interrupt injection formalization}
  \label{fig:irqc_sys_arch}
\end{figure}

%We define $P_j_{VM$ as the $j_{th}$ parameter that influences the  behavior from the safety point of view (e.g., number of cache misses). At each instance, the values of parameters indicate the current status of the target VM. The interference parameters can be used to define the interference effect of the target VM, referred to as $e_{VM_i$. As shown in Equation XYZ, such effect is calculated by combining the interferences that affect the target VM and each interference is due to the corresponding interference parameter.

%\begin{equation}
%    \begin{split}
%        e_{VM_i} = & f_i(I_{0_{VM_i}}(\hat{P}_{0_{VM_i}}), \\
%         I_{1_{VM_i}}(\hat{P}_{1_{VM_i}}), &..., I_{N_{VM_i}}(\hat{P}_{N_{VM_i}}),
%        )
%    \end{split}
%\end{equation}

%\begin{figure}[ht]
%  \centering
%  \includegraphics[width=\columnwidth]{images/IRQ Coloring/formulization_error.pdf}
%  \caption{Mibench benchmark automotive suite profiling - each benchmark is profiled in ten different periods}
%  \label{fig:irqc_sys_arch}
%\end{figure}

\subsection{IRQ Coloring RTM}   %%%%%%%%%%%%%%%%%%%%%%%%%%%%%%%%%%%%%%%%%%%%%%%%%%%%%%%%%%%%%%%%%%%%%%%%%%%%%%%%%%%%%%%%%%%%%%%%%%%%%%%%%%%%%%%%%%%%%%%%%%%%%%%%%%%%%%%%%%%%%%%%%%%%%%%%%%%%%%%%%%%%%%%%%

The IRQ Coloring RTM introduces the concept of \ac{QoS} awareness, i.e., the performance of each \ac{VM} will be evaluated and used to determine the next degradation mode to be applied to the system. To achieve this, operation tables will be created at design time using Karnaugh maps. These tables will enable the DTT to generate the control equation that the hypervisor will use to select the next stage to be applied. To ensure a modular design that can be easily modified and adapted for different setups, we have defined three stages of operation:

\begin{itemize}
\item \textbf{Stage 0} - Calculate the QoS for each \ac{VM}
\item \textbf{Stage 1} - Decode the QoS for each \ac{VM}
\item \textbf{Stage 2} - Calculate the next degradation mode
\end{itemize}

The IRQ coloring technique is applied based on the principle that the hypervisor is interrupted periodically at a fixed time interval. During this synchronous event, each \ac{VM} calculates its performance. The high-criticality \ac{VM} (e.g., \ac{ASIL}-D)   then uses this performance data to determine the new degradation mode to be implemented in the system. A detailed description of the process is provided in Algorithm \ref{alg:irq_col}.

\begin{algorithm}[b!]
\caption{IRQ Coloring RTM}
\label{alg:irq_col}
\begin{algorithmic}[1]
\Function{RTM timer handler}{}
\For{$i=1$ to $M$}
\If{$SIL(i) \neq QM$}
\State $qos \gets$ ComputeQoS$(\hat{P})$
\State $cf \gets$ DecodeQoS$(qos)$
\If{$SIL(i) = D$}
\State $deg\_mode \gets$ ComputeDM$(cf)$
\State MaskIRQs$(deg\_mode)$
\EndIf
\EndIf
\EndFor
\State RescheduleTimer
\EndFunction
\end{algorithmic}
\end{algorithm}

\mypara{Stage 0 - ComputeQoS().} In order to determine the performance of each \ac{VM}, we leveraged a weighted average of microarchitectural events $\hat{P}$, such as the number of cache access and bus access cycles, as the function $f_i(.)$ (different optimization techniques can be used). This computation takes into account the ratio between the actual number of accesses and the expected number of accesses, which is the number of accesses that would occur without interference during a given time interval. The reference values for these events and the weights used in the calculation of the weighted average should be defined by the DTT. The output of this computation, referred to as $qos$, is a value between 0 and 100 that represents the performance of the \ac{VM}.

\mypara{Stage 1 - DecodeQoS().} Once the \ac{QoS} value of a \ac{VM} has been calculated, it needs to be mapped to a representation that can be used in the control equation. This is achieved by applying the following mapping criteria:

\begin{equation} \label{eu_eqn}
\begin{aligned} 
    \forall \text{qos} \in ]100, 75], cf &= T0  \\
    \forall \text{qos} \in ]75, 50], cf &= T1   \\
    \forall \text{qos} \in ]50, 25], cf &= T2   \\
    \forall \text{qos} \in ]25, 0], cf &= T3
\end{aligned}
\end{equation}

This method allows the definition of a 2-bit representation $cf$ of the \ac{QoS} value for a specific \ac{VM}. The global control register for the IRQ Coloring RTM is the result of the aggregation of control flags associated with each \ac{VM}.

\mypara{Stage 2 - ComputeDM().} The computation of the next degradation mode - $deg\_mode$ - relies on the global control register calculated by the different \ac{VM}s and the control equation implemented at the hypervisor level. In Stages 0 and 1, all \ac{VM}s compute the RTM logic in parallel, but not in Stage 2. Since the calculations in Stage 2 rely on the previous stages, performing the very same computation per \ac{VM} is unnecessary. Instead, the higher criticality \ac{VM} calculates the next degradation mode, which is then applied to the system. %Once the computation is done, all \ac{VM}s mask the corresponding IRQs by manipulating the $ISENABLER$ and $ICENABLER$ registers from GIC to set and clear interrupts.

\begin{figure*}[ht]
  \centering
  \includegraphics[width=0.95\textwidth]{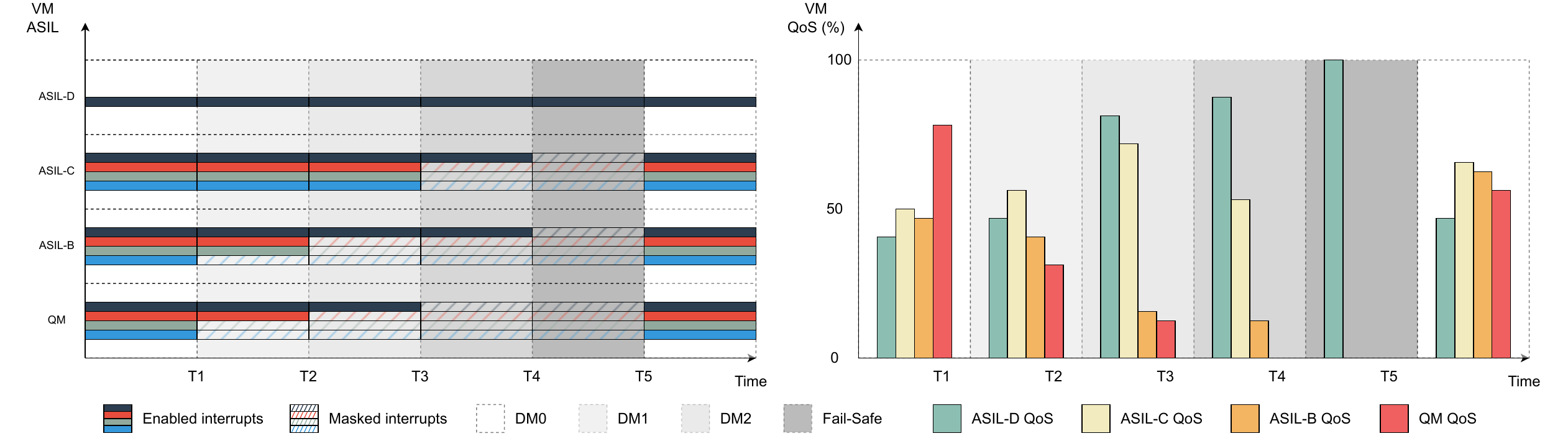}
  \caption{IRQ Coloring (RTM) "Toy" example. On the left, we depict the evolution of masked interrupts at different time intervals for all VMs. Each color represents a different interrupt, while the hatched interrupts correspond to the masked ones. On the right, we present each VM's resulting \ac{QoS} after masking the interrupts (identified on the left).}
  \label{fig:irqc_toy_ex}
\end{figure*}

\subsection{IRQ Coloring "Toy" Example} 

Figure \ref{fig:irqc_toy_ex} provides an illustrative example of the IRQ Coloring inner works. The setup comprises four \ac{VM}s, with one representing a highly critical system (ASIL-D), two others simulating intermediate critical systems (ASIL-C and ASIL-B), and a non-critical system (QM). At the initial instant (T0-T1), four interrupts are assigned to each intermediate and lower criticality \ac{VM}, and one interrupt is assigned to the higher criticality system, all of which are active.

After a given time interval, the hypervisor collects a series of events ($\hat{P}$) that enable the computation of the \ac{QoS}  of each \ac{VM}, which allows calculating the effect of interference ($e_{VM_i}$). Based on this value, the degradation mode is updated, masking two interruptions of the lower criticality \ac{VM}. However, applying this degradation mode alone is insufficient to guarantee the requirements of the most critical \ac{VM}s. Thus, the degradation mode is updated after another time period(T2). This process repeats until a fail-safe mode is reached (T4-T5), which involves suspending all non-critical \ac{VM}s. Once the requirements of the most critical \ac{VM} (ASIL-D) are satisfied, the interrupts of the medium and low critical \ac{VM}s become active again. However, it's important to note that this process must be carried out progressively, as explained earlier.

%%%%%%%%%%%%%%%%%%%%%%%%%%%%%%%%%%%%%%%%%%%%%%%%%%%%%%%%%%%%%%%%%%%%%%%%%%%%%%%%%%%%%%%%%%%%%%%%%%%%%%%%%%%%%%%%%%%%%%%%%%%%%%%%%%%%%%%%%%%%%%%%%%%%%%%%%%%%%%%%%

% The DTT aims to create all the necessary artifacts for the RTM by following a five-step process, which is illustrated in Figure \ref{fig:DTT_overview}. The first step involves the configuration of the overall setup \circled{1}, including the number of VMs and their characteristics. The second step involves the generation of the default configuration map (degradation modes) \circled{2}. The third step involves the manual tuning and updating of the default configuration map \circled{3}. The fourth step involves feeding the configuration map to the tool to construct RTM artifacts \circled{4}. Finally, the fifth step involves generating the hypervisor artifacts needed to implement the IRQ RTM \circled{5}. The tool is implemented in Python and can be configured through two YAML files and a spreadsheet, which makes it easy to use.

\section{Implementation} \label{sec:implementation}

\mypara{Hardware Implementation} We implemented and deployed the IRQ Coloring RTM on a Xilinx ZCU102 board featuring an Ultrascale+ ZU7EV SoC. The system has a quad-core Arm Cortex-A53 processor operating at 1.2GHz. CPUs have separate 32KiB L1 instruction and data caches and a shared 1MiB L2 cache. The cluster features the GIC-400 (GICv2). Nevertheless, it is possible to port the IRQ Coloring to a newer version of GIC, such as GICv3, since most of the IRQ Coloring RTM's functionality relies on the GICD, which is also available in GICv3. However, some changes may be necessary as although the $ISENABLER$ and $ICENABLER$ registers exist in both GICv2 and GICv3, there are subtle differences in their register layout and functionality.

%One crucial factor that determines the effectiveness of interference mitigation and intermediate guarantees of the IRQ Coloring technique is the system configuration and the overall performance overhead. The \ac{IRQ} Coloring \ac{RTM} was designed with this constraint in mind. To ensure that the \ac{IRQ} Coloring mechanism does not significantly degrade system performance, an optimization was considered in the form of using an \ac{OCM} for data storage. This would prevent data from being cached, thereby avoiding cache misses and evictions, which could lead to non-deterministic execution times of the mechanism and potentially harm the performance of the \ac{VM}. However, it was found that the benefits of using the \ac{OCM} were not as significant as expected, as the amount of data used by the \ac{IRQ} Coloring \ac{RTM} was relatively low. Therefore, it is optional whether to place the data structures on the \ac{OCM} or not.

\mypara{Software Implementation} The implementation of the \ac{IRQ} Coloring \ac{RTM} consists of a thin layer of software that sits between the distributor and the \ac{CPU} interfaces, built on top of the Bao hypervisor. The \ac{RTM} uses two hardware components to track the performance of \ac{VM}s and employs a time-based control system. The first component is the \ac{PMU} of each \ac{CPU}, which collects specific metrics to assess the \ac{QoS} of each \ac{VM}. The second component is the Generic Timer, which provides a time reference for the system and controls interrupt masking based on the established \ac{IRQ} Coloring policy. To implement the IRQ Coloring, we assume a 1-1 mapping of virtual to physical CPUs. Therefore, each \ac{PMU} collects the microarchitectural events of a single \ac{VM}. \ac{PMU} events, accessed via the MRS and MSR instructions, are used to compute the stages of the IRQ Coloring RTM. Once the computation is complete, all \ac{VM}s mask the corresponding \ac{IRQ}s by manipulating the $ISENABLER$ and $ICENABLER$ registers from the GIC to set and clear interrupts.

The configurability of the \ac{IRQ} Coloring \ac{RTM} is crucial for its effectiveness, adaptability, and scalability. Thus, it is key to provide an easy-to-configure interface; however, the configuration process is not limited to the interface, as the \ac{IRQ} Coloring masking policy also plays a vital role in ensuring proper distribution of interrupts among the \ac{VM}s. The masking policy must be carefully configured, taking into account the varying performance requirements of the \ac{VM}s, to ensure that critical interrupts are not delayed or lost. Moreover, the \ac{IRQ} Coloring mechanism offers configurability in several hyperparameters. These include (i) the actuation period, which determines the frequency of actuation of the mechanism, (ii) the weight assigned to each PMU event, and (iii) the expected behavior of each VM (e.g., the optimal number of cache misses that match the native execution of the VM). While the fine-tuning of these parameters can further enhance the mechanism's effectiveness in mitigating interference, the incorrect definition of these parameters, on the other hand, can lead to an incorrect actuation of the IRQ Coloring RTM. In order to refine the IRQ Coloring parameters, it is imperative to use a profiling tool that (i) enables the selection of microarchitectural events that better reflects the characteristics of the applications under consideration, (ii) as well as suggests the appropriate value for each reference parameter.

%\clearpage
\section{Evaluation} \label{sec:evaluation}

In this section we describe the evaluation setup and present and discuss the evaluation results.

\begin{table*}
\caption{Setup configurations and masked interrupts for each degradation mode (three different setup configurations). Each row represents a different degradation mode, and each column corresponds to the interrupt state for a specific setup configuration.}
\label{tab:masking_config}
\resizebox{\linewidth}{!}{%
\begin{tabular}{|c|l|c|c|c|l|l|c|c|c|c|c|c|c|l|} 
\cline{3-15}
\multicolumn{1}{c}{}                                                      &                      & ASIL-D                & \multicolumn{4}{c|}{ASIL-C}                                                                                                                               & \multicolumn{4}{c|}{ASIL-B}                                                                                                                                              & \multicolumn{4}{c|}{QM}                                                                                                                                    \\ 
\cline{3-15}
\multicolumn{1}{c}{}                                                      &                      & IRQ                   & IRQ 0                                & IRQ 1                                & IRQ 2                                & IRQ 3                                & IRQ 0                                & IRQ 1                                & IRQ 2                                & IRQ 3                                               & IRQ 0                                & IRQ 1                                & IRQ 2                                & IRQ 3                                 \\ 
\hline
\multirow{4}{*}{Dual-VM}                                                  & Degradation Mode 0   &                       & {\cellcolor[rgb]{0.8,0.8,0.8}}       & {\cellcolor[rgb]{0.8,0.8,0.8}}       & {\cellcolor[rgb]{0.8,0.8,0.8}}       & {\cellcolor[rgb]{0.8,0.8,0.8}}       & {\cellcolor[rgb]{0.8,0.8,0.8}}       & {\cellcolor[rgb]{0.8,0.8,0.8}}       & {\cellcolor[rgb]{0.8,0.8,0.8}}       & {\cellcolor[rgb]{0.8,0.8,0.8}}                      &                                      &                                      &                                      & {\cellcolor[rgb]{0.502,0.502,0.502}}  \\ 
\hhline{|~--------------|}
                                                                          & Degradation Mode 1   &                       & {\cellcolor[rgb]{0.8,0.8,0.8}}       & {\cellcolor[rgb]{0.8,0.8,0.8}}       & {\cellcolor[rgb]{0.8,0.8,0.8}}       & {\cellcolor[rgb]{0.8,0.8,0.8}}       & {\cellcolor[rgb]{0.8,0.8,0.8}}       & {\cellcolor[rgb]{0.8,0.8,0.8}}       & {\cellcolor[rgb]{0.8,0.8,0.8}}       & {\cellcolor[rgb]{0.8,0.8,0.8}}                      &                                      &                                      & {\cellcolor[rgb]{0.502,0.502,0.502}} & {\cellcolor[rgb]{0.502,0.502,0.502}}  \\ 
\hhline{|~--------------|}
                                                                          & Degradation Mode 2   &                       & {\cellcolor[rgb]{0.8,0.8,0.8}}       & {\cellcolor[rgb]{0.8,0.8,0.8}}       & {\cellcolor[rgb]{0.8,0.8,0.8}}       & {\cellcolor[rgb]{0.8,0.8,0.8}}       & {\cellcolor[rgb]{0.8,0.8,0.8}}       & {\cellcolor[rgb]{0.8,0.8,0.8}}       & {\cellcolor[rgb]{0.8,0.8,0.8}}       & {\cellcolor[rgb]{0.8,0.8,0.8}}                      &                                      & {\cellcolor[rgb]{0.502,0.502,0.502}} & {\cellcolor[rgb]{0.502,0.502,0.502}} & {\cellcolor[rgb]{0.502,0.502,0.502}}  \\ 
\hhline{|~--------------|}
                                                                          & Fail-Safe Strategy   &                       & {\cellcolor[rgb]{0.8,0.8,0.8}}       & {\cellcolor[rgb]{0.8,0.8,0.8}}       & {\cellcolor[rgb]{0.8,0.8,0.8}}       & {\cellcolor[rgb]{0.8,0.8,0.8}}       & {\cellcolor[rgb]{0.8,0.8,0.8}}       & {\cellcolor[rgb]{0.8,0.8,0.8}}       & {\cellcolor[rgb]{0.8,0.8,0.8}}       & {\cellcolor[rgb]{0.8,0.8,0.8}}                      & {\cellcolor[rgb]{0.502,0.502,0.502}} & {\cellcolor[rgb]{0.502,0.502,0.502}} & {\cellcolor[rgb]{0.502,0.502,0.502}} & {\cellcolor[rgb]{0.502,0.502,0.502}}  \\ 
\hline
\multirow{4}{*}{\begin{tabular}[c]{@{}c@{}}Quad-VM\\Setup 1\end{tabular}} & Degradation Mode 0   &                       &                                      &                                      & {\cellcolor[rgb]{0.8,0.8,0.8}}       & {\cellcolor[rgb]{0.8,0.8,0.8}}       &                                      &                                      &                                      &                                                     &                                      &                                      &                                      &                                       \\ 
\hhline{|~--------------|}
                                                                          & Degradation Mode 1   &                       &                                      &                                      & {\cellcolor[rgb]{0.8,0.8,0.8}}       & {\cellcolor[rgb]{0.8,0.8,0.8}}       &                                      &                                      & {\cellcolor[rgb]{0.502,0.502,0.502}} & {\cellcolor[rgb]{0.502,0.502,0.502}}                &                                      &                                      & {\cellcolor[rgb]{0.502,0.502,0.502}} & {\cellcolor[rgb]{0.502,0.502,0.502}}  \\ 
\hhline{|~--------------|}
                                                                          & Degradation Mode 2   &                       &                                      & {\cellcolor[rgb]{0.502,0.502,0.502}} & {\cellcolor[rgb]{0.8,0.8,0.8}}       & {\cellcolor[rgb]{0.8,0.8,0.8}}       &                                      & {\cellcolor[rgb]{0.502,0.502,0.502}} & {\cellcolor[rgb]{0.502,0.502,0.502}} & {\cellcolor[rgb]{0.502,0.502,0.502}}                & {\cellcolor[rgb]{0.502,0.502,0.502}} & {\cellcolor[rgb]{0.502,0.502,0.502}} & {\cellcolor[rgb]{0.502,0.502,0.502}} & {\cellcolor[rgb]{0.502,0.502,0.502}}  \\ 
\hhline{|~--------------|}
                                                                          & Fail-Safe Strategy   &                       & {\cellcolor[rgb]{0.502,0.502,0.502}} & {\cellcolor[rgb]{0.502,0.502,0.502}} & {\cellcolor[rgb]{0.8,0.8,0.8}}       & {\cellcolor[rgb]{0.8,0.8,0.8}}       & {\cellcolor[rgb]{0.502,0.502,0.502}} & {\cellcolor[rgb]{0.502,0.502,0.502}} & {\cellcolor[rgb]{0.502,0.502,0.502}} & {\cellcolor[rgb]{0.502,0.502,0.502}}                & {\cellcolor[rgb]{0.502,0.502,0.502}} & {\cellcolor[rgb]{0.502,0.502,0.502}} & {\cellcolor[rgb]{0.502,0.502,0.502}} & {\cellcolor[rgb]{0.502,0.502,0.502}}  \\ 
\hline
\multirow{4}{*}{\begin{tabular}[c]{@{}c@{}}Quad-VM\\Setup 2\end{tabular}} & Degradation Mode 0   &                       &                                      &                                      &                                      &                                      &                                      &                                      &                                      &                                                     &                                      &                                      &                                      &                                       \\ 
\hhline{|~--------------|}
                                                                          & Degradation Mode 1   &                       &                                      &                                      &                                      & {\cellcolor[rgb]{0.502,0.502,0.502}} &                                      &                                      &                                      & {\cellcolor[rgb]{0.502,0.502,0.502}}                &                                      &                                      & {\cellcolor[rgb]{0.502,0.502,0.502}} & {\cellcolor[rgb]{0.502,0.502,0.502}}  \\ 
\hhline{|~--------------|}
                                                                          & Degradation Mode 2   &                       &                                      &                                      & {\cellcolor[rgb]{0.502,0.502,0.502}} & {\cellcolor[rgb]{0.502,0.502,0.502}} &                                      & {\cellcolor[rgb]{0.502,0.502,0.502}} & {\cellcolor[rgb]{0.502,0.502,0.502}} & {\cellcolor[rgb]{0.502,0.502,0.502}}                & {\cellcolor[rgb]{0.502,0.502,0.502}} & {\cellcolor[rgb]{0.502,0.502,0.502}} & {\cellcolor[rgb]{0.502,0.502,0.502}} & {\cellcolor[rgb]{0.502,0.502,0.502}}  \\ 
\hhline{|~--------------|}
                                                                          & Fail-Safe Strategy   &                       & {\cellcolor[rgb]{0.502,0.502,0.502}} & {\cellcolor[rgb]{0.502,0.502,0.502}} & {\cellcolor[rgb]{0.502,0.502,0.502}} & {\cellcolor[rgb]{0.502,0.502,0.502}} & {\cellcolor[rgb]{0.502,0.502,0.502}} & {\cellcolor[rgb]{0.502,0.502,0.502}} & {\cellcolor[rgb]{0.502,0.502,0.502}} & {\cellcolor[rgb]{0.502,0.502,0.502}}                & {\cellcolor[rgb]{0.502,0.502,0.502}} & {\cellcolor[rgb]{0.502,0.502,0.502}} & {\cellcolor[rgb]{0.502,0.502,0.502}} & {\cellcolor[rgb]{0.502,0.502,0.502}}  \\ 
\hline
\multicolumn{1}{c}{}                                                      & \multicolumn{1}{l}{} & \multicolumn{1}{l}{}  & \multicolumn{1}{l}{}                 & \multicolumn{1}{l}{}                 & \multicolumn{1}{l}{}                 & \multicolumn{1}{l}{}                 & \multicolumn{1}{l}{}                 & \multicolumn{1}{l}{}                 & \multicolumn{1}{l}{}                 & \multicolumn{1}{l}{}                                & \multicolumn{1}{l}{}                 & \multicolumn{1}{l}{}                 & \multicolumn{1}{l}{}                 & \multicolumn{1}{l}{}                  \\ 
\hhline{~~-~~~-~~~-~~~~}
\multicolumn{1}{c}{}                                                      &                      & \multicolumn{1}{l|}{} & \multicolumn{2}{l}{Enabled IRQ}                                             & \multicolumn{1}{c|}{}                & {\cellcolor[rgb]{0.502,0.502,0.502}} & \multicolumn{2}{l}{Masked IRQ}                                              & \multicolumn{1}{l|}{}                & \multicolumn{1}{l|}{{\cellcolor[rgb]{0.8,0.8,0.8}}} & \multicolumn{2}{l}{N.A. IRQ}                                                & \multicolumn{1}{l}{}                 & \multicolumn{1}{l}{}                  \\
\hhline{~~-~~~-~~~-~~~~}
\end{tabular}
}
\end{table*}

\subsection{Methodology}

%\mypara{Hardware Platform.} Tests were conducted using a Zynq Ultrascale+ ZU7EV system-on-chip on a Xilinx ZCU102 evaluation board. The system is equipped with a quad-core Arm Cortex-A53 processor, operating at 1.2GHz. The individual CPUs are equipped with separate 32KiB L1 instruction and data caches, and a shared 1MiB L2 cache. The cluster features the GIC-400 (GICv2).

\mypara{Evaluation Setup.} To assess the effectiveness of the IRQ Coloring RTM, the evaluation comprises two different use cases. The first use case involves two \ac{VM}, each assigned one CPU. The first \ac{VM} runs on a Linux OS and executes various suites of the Mibench benchmark. The second \ac{VM} is intended for memory-intensive applications and aims to create contention at the last-level cache and system bus. This \ac{VM} comprises four tasks, each triggered by a different interrupt generated by a hardware module deployed on the PL of the ZCU102. The second use case expands the evaluation to four \ac{VM} - 1 CPU assigned to each VM. The critical VM runs a synthetic application that mimics the behavior of the MiBench Benchmark\footnote{The evaluation was conducted targeting an (emulated) automotive use case, using Erika3 to set up the evaluation environment. At the time of this writing, we could not find a fully functional Mibench AICS Suite port for Erika3 RTOS. Thus, we could not use the original Mibench AICS Suite, which is ready-to-use for Linux. Therefore, an adaptation of the benchmark was required to make it compatible with Erika3. To achieve this, we performed a benchmark profiling process by collecting microarchitectural events, such as L2 cache accesses and bus accesses, during the execution of each benchmark. Based on this profiling, we reconstructed the benchmark so that the application would perform load and store operations to replicate the behavior of the original benchmark. We called this suite the "Synthetic Mibench AICS Suite".}, while the other three run memory-intensive applications. The objective is to measure the impact of increased number of irq-driven workloads on the performance of the higher criticality VMs and, in parallel, understand the intermediate guarantees of the IRQ Coloring RTM mechanism on the medium criticality VMs.

\mypara{VM Workload.} In our assessment of the IRQ Coloring RTM, we employed the well-established MiBench Automotive and Industrial Control System (AICS) Suite in the critical VM. This subset includes three of the most memory-intensive benchmarks, namely qsort, susan corners, and susan edges (Figure \ref{fig:mibench_profile}). We chose these benchmarks based on the observed profiling, i.e., these benchmarks are more susceptible to interference arising from cache and memory contention. In order to create interference among the \ac{VM} (VMs), we deployed a baremetal application that continuously writes into a buffer equivalent in size to the last-level cache (LLC) (i.e., 1 MiB). Each VM is assigned four distinct interrupts, each of which activates a workload that writes to different segments of the buffer. To assess the impact of different workloads on the system's performance, we divided the buffer into four sections: a 512KiB partition (equivalent to 50\% of the LLC), a 256KiB partition (equivalent to 25\% of the LLC), and two partitions of 128KiB each (equivalent to 12.5\% of the LLC).

\mypara{Measurement Tools.} We use the Arm \ac{PMU} to collect microarchitectural events to profile the benchmark execution. The chosen events comprise the number of L2 cache accesses and the number of bus accesses. Additionally, we leverage the "perf" tool on the Linux operating system to measure the execution time of each benchmark.

\mypara{IRQ Coloring RTM Configuration.} To ensure the optimal performance of the IRQ Coloring RTM in mitigating interrupt interference, we made the following assumptions during our evaluation:

\begin{enumerate}
    \item The weights assigned to each microarchitectural event are equally distributed (50/50).
    \item The IRQ Coloring RTM is sampled at a frequency of 10\% of the benchmark execution time to avoid the risk of aliasing due to the Nyquist-Shannon sampling theorem \cite{nyquist1928certain, shannon1949communication} while capturing sufficient information.
    \item The interrupt masking on intermediate and non-critical VMs is based on the profile of each task, and the masking maps applied to each setup used in the Evaluation section are presented in Table \ref{tab:masking_config}.
\end{enumerate}

\subsection{IRQ Coloring RTM Microbenchmark}

\begin{figure}[t!]
  \centering
  \includegraphics[width=\linewidth]{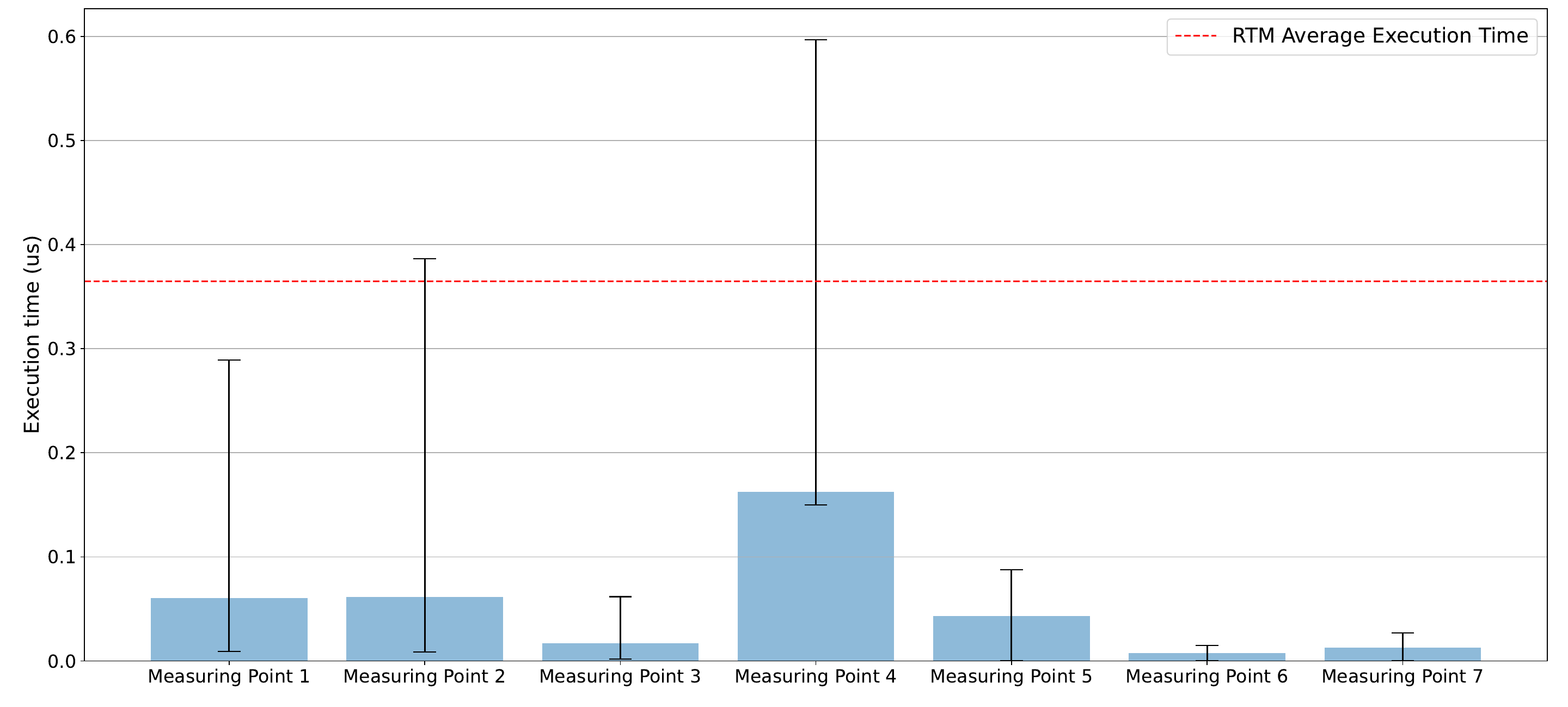}
  \caption{Microbenchmark results for the IRQ Coloring RTM: average and worst-case execution times for the different measuring points identified in Table \ref{tab:rtm_microbenchmark}.}
  \label{fig:rtm_microbm}
\end{figure}

\begin{figure}[t!]
  \centering
  \includegraphics[width=0.85\linewidth]{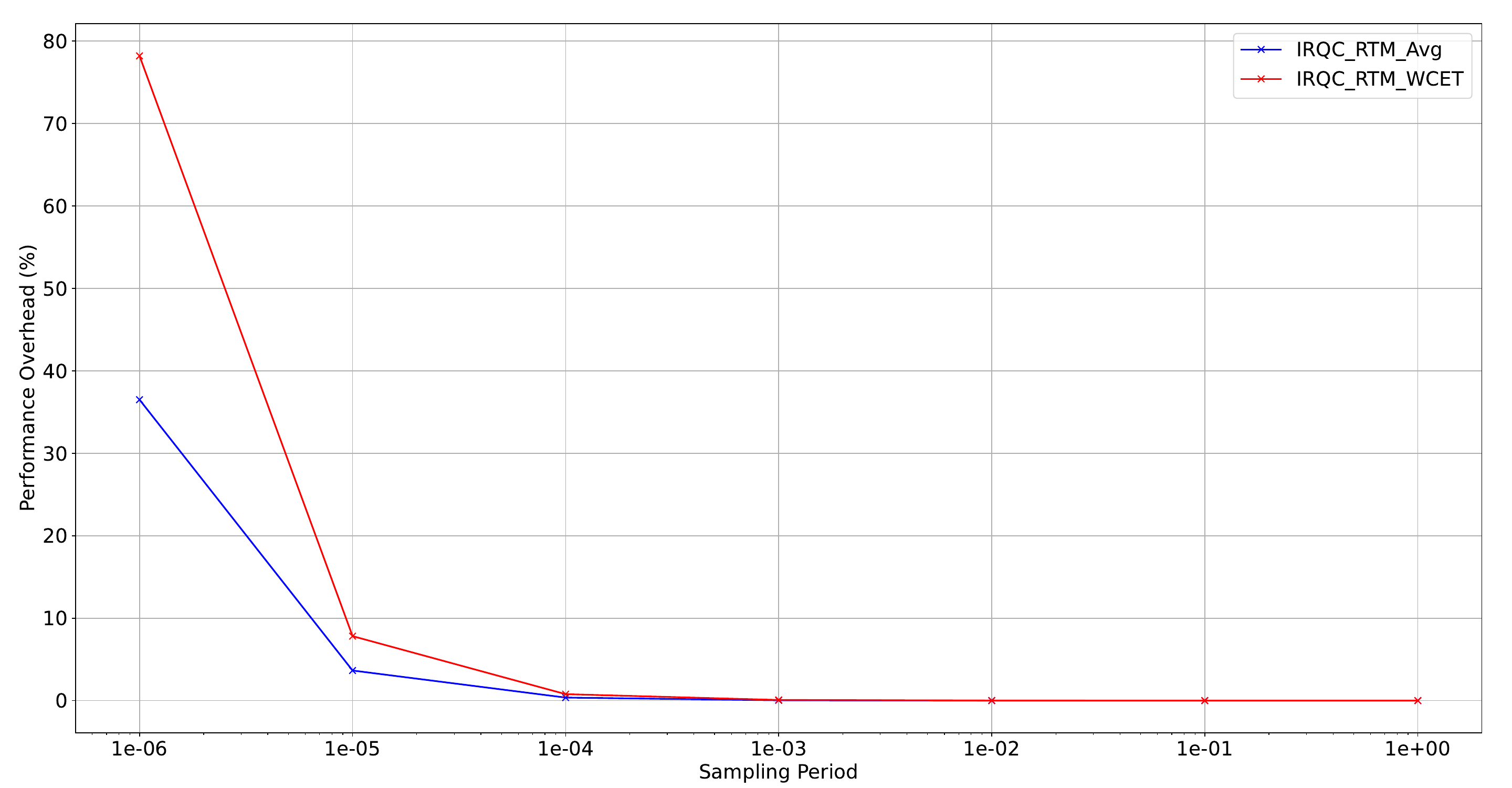}
  \caption{IRQ Coloring RTM: performance overhead vs. actuation period}
  \label{fig:rtm_overhead}
\end{figure}

Figure \ref{fig:rtm_microbm} illustrates the IRQ Coloring RTM microbenchmark, which comprises the seven measuring points detailed in Table \ref{tab:rtm_microbenchmark}. The first measuring point corresponds to the "PMU Sampling", which represents the time taken by the processor to collect performance data using the \ac{PMU}. The next two measuring points correspond to different stages of the mechanism, with Stage 0 encompassing the QoS computation for each VM and Stage 1 corresponding to the decoding of the QoS value. The next three measuring points correspond to Stage 2, which involves different VMs synchronization, reference update, and interrupt masking operations. These stages are key for the IRQ Coloring RTM overall performance, and any inefficiencies or delays at this stage can significantly impact the overall system's performance. The profiling results of the IRQ Coloring RTM indicate that the mechanism's average execution time is 0.365 microseconds. However, due to the need to synchronize the different VMs, the maximum execution time observed is 0.782 microseconds. To put this into perspective, if the actuation period is 10 microseconds, the expected performance impact is 7.81\%. In fact, the performance impact becomes negligible for sampling periods above 100 microseconds, resulting in a performance impact value below 0.8\%, as depicted in Figure \ref{fig:rtm_overhead}. Therefore, sampling the system at a frequency of 10 kHz would not have a significant performance impact.

\begin{figure*}[t!]
  \centering
  \includegraphics[width=0.99\textwidth]{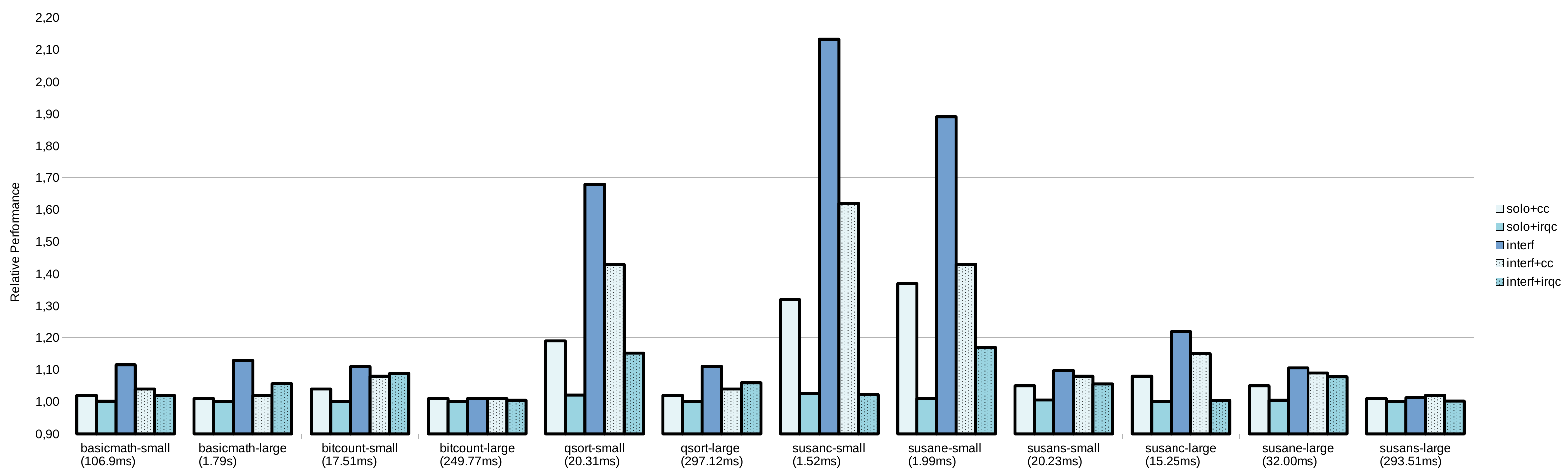}
  \caption{Dual-VM: performance degradation for the Mibench AICS under interference, cache coloring, and IRQ coloring}
  \label{fig:2-core-results}
  \vspace{-0.1cm}
\end{figure*}

\begin{table}
\centering
\caption{Description of micro-operations (measuring points) of the IRQ Coloring RTM}
\label{tab:rtm_microbenchmark}
\resizebox{\linewidth}{!}{%
\begin{tabular}{|c|l|} 
\hline
\begin{tabular}[c]{@{}c@{}}\textbf{Measuring }\\\textbf{Point}\end{tabular} & \multicolumn{1}{c|}{\textbf{Description}}                                                                                                                  \\ 
\hline
1                                                                           & Time taken by the RTM for the PMU sampling.                                                                                                                \\ 
\hline
2                                                                           & \begin{tabular}[c]{@{}l@{}}Time taken by the RTM to calculate the QoS for each VM \\based on the collected PMU values (from previous stage).\end{tabular}  \\ 
\hline
3                                                                           & Time taken by the RTM to decode the VM QoS values.                                                                                                         \\ 
\hline
4                                                                           & \begin{tabular}[c]{@{}l@{}}Time taken by the RTM to synchronize all VMs before \\processing / selecting the next degradation mode.\end{tabular}            \\ 
\hline
5                                                                           & \begin{tabular}[c]{@{}l@{}}Time taken by the RTM to run the control logic responsible\\for computing the next degradation mode.\end{tabular}               \\ 
\hline
6                                                                           & \begin{tabular}[c]{@{}l@{}}Time taken by the RTM to update reference values for the \\next degradation mode.\end{tabular}                                  \\ 
\hline
7                                                                           & \begin{tabular}[c]{@{}l@{}}Time taken by the RTM for interrupt masking operation per \\the mapping for the next degradation mode.\end{tabular}             \\
\hline
\end{tabular}
}
\end{table}

\subsection{Results with Mibench Benchmaks - Dual-VM}

We start by assessing the impact of interference mitigation on the high-criticality \ac{VM}. The setup consists of two VMs running atop of hypervisor: (i) a Linux-based VM running the MiBench AICS Suite and (ii) a low-criticality (e.g., ASIL-QM) running a synthetic benchmark to create contention at the \ac{LLC}. The evaluation process consists of comparing the execution of the MiBench AICS Suite benchmarks for six different system configurations: (i) solo benchmark execution (solo); (ii) solo benchmark execution with cache coloring enabled (solo+cc); (iii) solo benchmark execution with IRQ coloring enabled (solo+irqc); (iv) benchmark execution under interference (interf); (v) benchmark execution under interference with cache coloring enabled (interf+cc); and (vi) benchmark execution under interference with IRQ coloring enabled (interf+irqc).
%in standalone mode (solo) and under contention (interf) to assess the level of interference that occurs when running multiple guests concurrently. The evaluation is then repeated with cache partitioning enabled (solo-col and interf-col) to investigate the impact of the first level of micro-architectural partitioning on the target partitions and its potential to mitigate interference. Finally, the evaluation is reiterated with IRQ Coloring (solo-irqc and interf-irqc).

\mypara{Interference Mitigation Effect.} The average results for 1000 runs are presented in Figure \ref{fig:2-core-results}, and they are normalized to the native execution, where higher values indicate poorer performance. The empirical results demonstrate that memory-intensive applications, such as susan-c small, susan-e small, and q-sort small, can be significantly impacted by over 65\% due to contention arising from the sharing of microarchitectural resources like the LLC and system bus. In the worst-case scenario, which is susan-c small, the execution time increases from 1.53ms to 3.27ms (2.13x when compared to the native execution). However, using interference mitigation techniques such as cache partitioning can help reduce the interference effect. By allocating 50\% of the LLC to each VM, cache partitioning can significantly reduce interference and decrease the relative performance overhead to 1.62x for susan-c small, which is the most memory-intensive application, reducing the execution time to 2.48ms.

Furthermore, empirical results show that IRQ coloring is generally more effective in mitigating interference than the cache coloring technique, especially when the interference is very high, e.g., susan-c small, where the performance degradation is reduced from 2.13x to 1.02x, reducing the execution time from 3.27ms to 1.57ms. However, there are three cases where IRQ coloring is not as effective as cache coloring, i.e., basicmath-large, bitcount-small, and qsort-large. We investigated and we concluded that it is related to two major factors: (i) the profile of the benchmark, i.e., no memory-intensive and small execution time (few milliseconds); (ii) the weights of the control equation, i.e., 50-50, where the real benchmark has considerable more cache accesses than bus accesses (90-10).

\mypara{Performance Impact.} Leveraging techniques such as cache coloring can effectively reduce interference on multicore platforms. However, it is important to note that the performance degradation resulting from employing cache coloring due to LLC fragmentation is not negligible. For the susan-c small benchmark using cache coloring incurs an impact of 1.32x, increasing the execution time from 1.53ms to 2.02ms. In contrast, the IRQ coloring technique for the solo case has a near-negligible overhead ($\leq$ 1.01x). Only for the qsort-small and susanc-small benchmarks is the overhead near 1.02x-1.03x, which in the case of the susanc-small results in an increase in the execution time from 1.53ms to 1.57ms. Overall, this results from the reduced execution time, as discussed in Section V.B.

\subsection{Results with Mibench Benchmaks - Quad-VM}

\begin{figure*}
  \centering
  \subfigure[Setup 1: ASIL-D (1 IRQ) and ASIL-C (2 IRQ) running on top of Erika3, ASIL-B (4 IRQ) and QM (4 IRQ) running on top of a baremetal]{\includegraphics[width=0.47\textwidth]{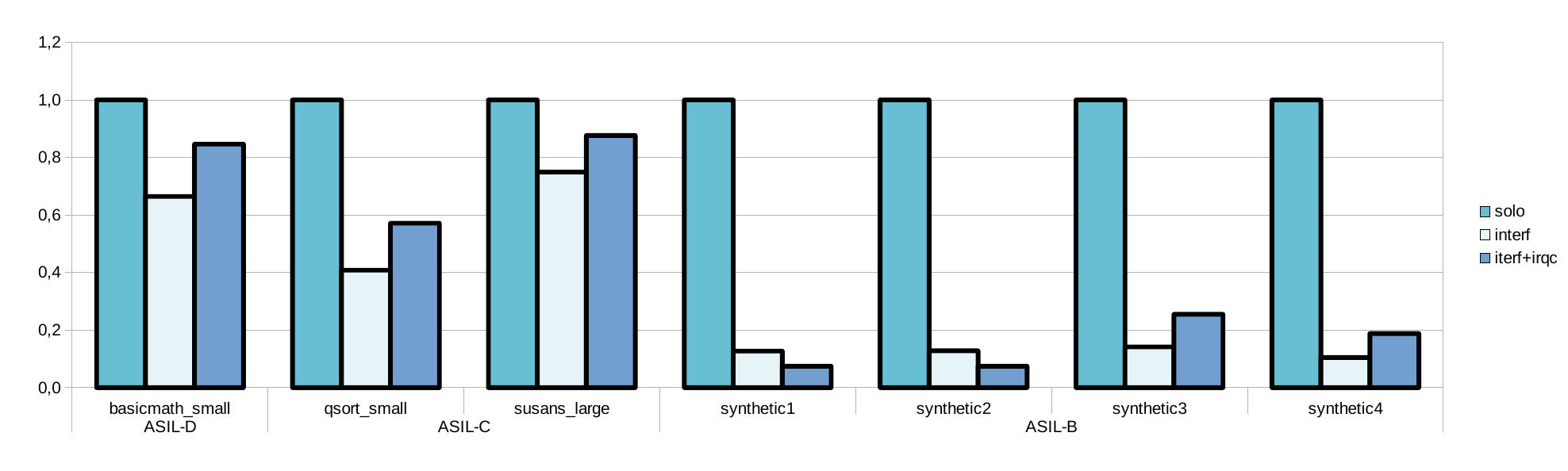}}\quad
  \subfigure[Setup 2: ASIL-D (1 IRQ) and ASIL-C (4 IRQ) running on top of Erika3, ASIL-B (4 IRQ) and QM (4 IRQ) running on top of a baremetal]{\includegraphics[width=0.47\textwidth]{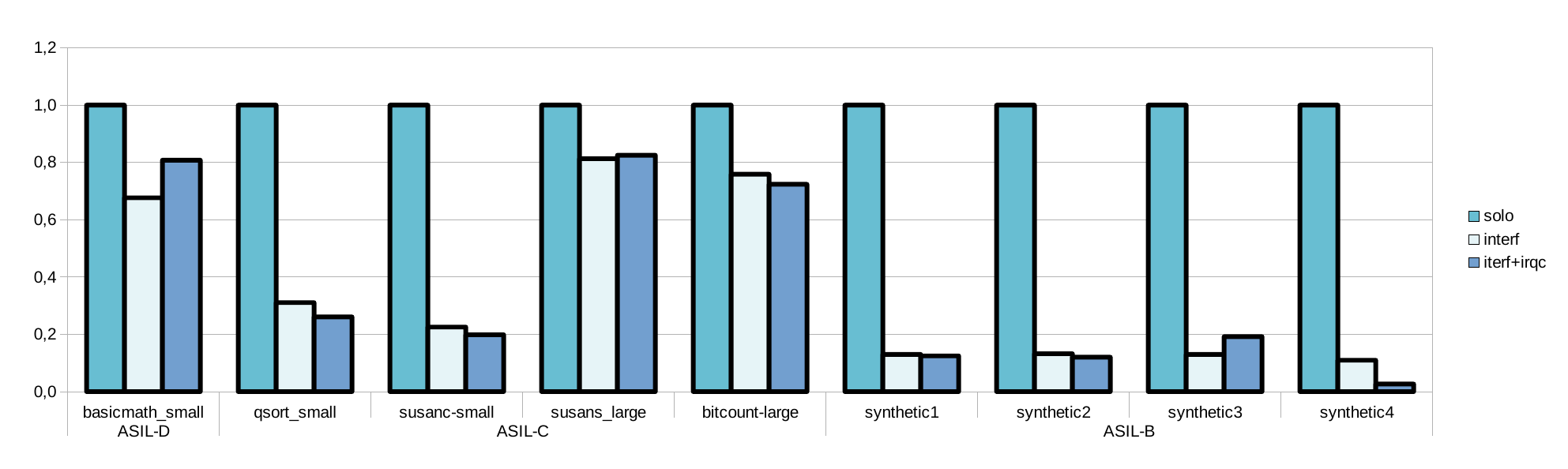}}\\
  %\subfigure[Setup 3: ASIL-D (1 IRQ), ASIL-C (2 IRQ) and ASIL-B (4 IRQ) running on top of Erika3, and QM (4 IRQ) running on top of a baremetal]{\includegraphics[width=0.47\textwidth]{images/results/setup4.pdf}}\quad
  %\subfigure[Setup 4: ASIL-D (1 IRQ), ASIL-C (4 IRQ) and ASIL-B (4 IRQ) running on top of Erika3, and QM (4 IRQ) running on top of a baremetal]{\includegraphics[width=0.47\textwidth]{images/results/setup5.pdf}}
  \caption{Assessment of intermediate guarantees in Quad-VM setup: comparison of relative throughput (y-axis) for each task across different VMs (ASIL) and benchmarks}
  \label{fig:quadVM_eval}
\end{figure*}

We will now focus on evaluating the intermediate guarantees offered by the IRQ Coloring mechanism. Many interference mitigation techniques on multicore platforms, such as memory throttling, rely on suspending a CPU, which halts its activity along with the corresponding VM. IRQ Coloring aims at providing intermediate guarantees to medium-critical VMs by gradually reducing their workload to minimize interference with higher-critical VMs. 
To gather evidence on the intermediate guarantees provided by IRQ Coloring, we deployed two different systems, each with four VMs. During a 10-second period, we ran the workloads of each VM and counted the number of times each task was completed. With that, we compute the availability of each VM based on the number of completed tasks. For each setup, we assessed the throughput of each VM by collecting their execution alone with the different active tasks (solo), their execution over interference with the four VMs (interf), and the interference scenario with IRQ Coloring (interf+irqc).

\mypara{Setup 1.} The first setup encompasses four VMs, each assigned to one CPU. This setup emulates a realistic automotive use case, with ASIL-D and ASIL-C VMs running Erika RTOS and automotive-related benchmarks. In this case, due to the absence of baremetal support from the original MiBench AICS Suite, we ran the synthetic MiBench Suite. The ASIL-B and QM VMs intended to emulate memory-intensive applications and create contention at the last-level cache and system bus. These VMs comprise four tasks, each triggered by a different interrupt generated by a hardware module deployed on the PL of the ZCU102. Figure \ref{fig:quadVM_eval}-(a) presents the workload bandwidth for the high-criticality VM, i.e., ASIL-D VM, and intermediate VMs, i.e., ASIL-C VM and ASIL-B VM. It is worth noting that no workload is completely stopped for both ASIL-C and ASIL-B VMs. For the ASIL-D and ASIL-C VMs, we highlight the additional bandwidth for both workloads (benchmarks) compared to the scenario under interference.

\mypara{Setup 2.} The second setup is an extension of Setup 1, designed to expand the workload of the ASIL-C VM. ASIL-D and ASIL-C VMs run Erika RTOS and the (synthetic) MiBench Suite. In this case, ASIL-C VM runs four benchmarks instead of two. The ASIL-B and QM VMs still consist of four tasks, each triggered by a different interrupt generated by the hardware module. Figure \ref{fig:quadVM_eval}-(b) presents the workload bandwidth for the high-criticality VM, i.e., ASIL-D VM, and intermediate VMs, i.e., ASIL-C VM and ASIL-B VM. It is worth noting that, compared to the results of Figure \ref{fig:quadVM_eval}-(a), the increased workload in the ASIL-C VM leads to a scenario where one of the benchmarks (susanc-small) presents less bandwidth than the scenario under interference.

%\mypara{Setup 3.} Setup 3 is a variant of Setup 1 with configuration changes related to the ASIL-B VM3, i.e., running another instance of Erika3 OS instead of baremetal. Compared to the results in Figure \ref{fig:quadVM_eval}-(a), adding an RTOS instance (Erika) for the ASIL-B VM has no noticeable effect on the overall bandwidth of the ASIL-D and ASIL-C VMs (results depicted in Figure \ref{fig:quadVM_eval}-(c)). However, it shows a significant deterioration in the bandwidth on the execution of the synthetic memory-intensive benchmarks on the ASIL-B VM.

%\mypara{Setup 4.} Finally, the fourth setup is a combination of Setup 2 and Setup 3. It features a higher workload on the ASIL-C VM (Setup 2) and changes in the configuration related to the workload of ASIL-B VM3, which runs an Erika OS instead of baremetal (Setup 3). Compared to the results of Figure \ref{fig:quadVM_eval}-(b), the addition of an RTOS instance (Erika) for the ASIL-B VM also shows a significant deterioration of the bandwidth on the execution of the synthetic memory-intensive benchmarks on the ASIL-B VM (results depicted in Figure \ref{fig:quadVM_eval}-(c)).

\section{Discussion}

%\textcolor{blue}{In this section, we describe a set of limitations and research directions regarding the IRQ Coloring DTT and RTM.}

%\textcolor{blue}{\mypara{Workload Profiling.} In this work, we profiled workloads by manually collecting execution time and microarchitectural events using the PMU. There are other profiling techniques available, such as static analysis of code, timing analysis via code instrumentation, and model-based analysis. These techniques can be explored further to profile applications. Additionally, the selection of key microarchitectural events to monitor can be automated, which is an avenue for future exploration.}

%\textcolor{blue}{\mypara{Configuration Map.} The IRQ Coloring DTT generates a preset configuration based on the system setup (e.g., number of VMs, number of IRQs). This preset configuration (also known as the default degradation modes map) translates the definition of the various degradation modes based on the performance of the different VMs. However, this preset configuration may be suboptimal for a given scenario, e.g.,  non-compliance with safety requirements or inadequate coverage of a specific setup. Fine-tuning the Configuration Map requires manual customization of the system designer. Another potential research direction would be to provide some intelligence (e.g., via machine learning models) that produces an optimized configuration map based on the automatic workload profile. }

\mypara{IRQ Coloring RTM QoS Calculation.} The RTM calculates the QoS of each VM by taking a weighted average of only two microarchitectural events: L2 cache accesses and bus access cycles. However, it is vital to understand the optimal weight of each event for better control logic, as observed in the evaluation of benchmarks. Adding other microarchitectural events to the control equation logic may impact the performance overhead and interference mitigation effectiveness; thus, it would be interesting to perform such a study in the near future. Furthermore, another potential optimization would be the use of a static look-up table calculated by the IRQ DTT instead of a weighted sum in the control logic.

\mypara{IRQ Coloring RTM QoS Reference Values.} The IRQ Coloring RTM QoS reference values are static and unique (single) per VM. However, workload behavior can change significantly during execution (as we observe while reconstructing the synthetic MiBench AICS Suite). To address this, having a vector of reference values per VM per execution period would be beneficial. This means having multiple reference values per QoS element per workload period; however, this poses additional challenges in synchronization. Investigating the impact of vector length on performance overhead and intermediate guarantees would also be interesting.

\mypara{IRQ Coloring RTM Actuation Period.} The IRQ Coloring RTM Actuation Period is crucial for performance overhead and interference mitigation effectiveness, which affects intermediate guarantees. To determine the actuation period, we considered the benchmark profile and interference mitigation effectiveness from the preliminary set of experiments. To be conservative, we set the actuation period ten times smaller than the execution period of the smallest application of the most critical VM, following the Nyquist-Shannon sampling theorem. However, this period may not be optimal for intermediate VMs when the high-criticality VM is idle. Dynamic set options for the RTM actuation period should be explored.

\mypara{Portability of IRQ Coloring.} IRQ Coloring is implemented for Armv8-A platforms. However, it can be adapted to other platforms, including new Arm real-time processors (Cortex-R52) and RISC-V application processors (e.g., CVA6). Evaluating the effectiveness of the technique on these platforms would be valuable, especially for Cortex-R52, which lacks hardware primitives for cache partitioning, and for RISC-V, which presents hardware-software co-design opportunities.

%\newpage
\section{Related work}

Several memory interference mitigation mechanisms have been put forth by the real-time research community. Those COTS-applicable mainly focus on shared LLC or DRAM bank partitioning, regulating memory bandwidth, or co-scheduling. Although mainly designed for OSs, these techniques have also been applied in hypervisors, but none leverage interrupt masking to achieve their end-goals.

\mypara{Cache Partitioning.} Cache partitioning \cite{Gracioli2015} consists of assigning subsets of the LLC to a specific workload and can be implemented in two main ways. Cache locking requires hardware assistance to restrict the eviction of selected cache lines, while cache coloring leverages virtual page number and cache index overlap to partition cache sets. \textit{Colored Lockdown} \cite{Mancuso2013} combines coloring and locking. Other works have proposed dynamic re-coloring schemes \cite{Ye2014, Xu2017, Roozkhosh2020}. Cache coloring has been implemented in several hypervisors such as Bao \cite{bao2020}, Jailhouse \cite{kloda2019deterministic}, and XVisor \cite{Modica2018}.

\mypara{DRAM Bank Partitioning} This technique leverages the parallelism in DRAM bank access to avoid contention among different workloads. PALLOC \cite{Yun2014} proposes an OS-level DRAM bank-aware memory allocator to avoid bank sharing among cores. In \cite{kloda2019deterministic}, authors combine DRAM bank and cache coloring into a single allocator at the hypervisor level. Similar ideas have been transposed to low-end microcontrollers \cite{pinto2019virtualization}.

\mypara{Memory Bandwidth Regulation}. By limiting the memory bandwidth of partitions, one can ensure bandwidth guarantees for higher-criticality workloads. Memguard \cite{MemGuard} throttles cores based on a memory bandwidth budget allocation, using performance monitoring counters to measure memory accesses and trigger the mechanism. In \cite{Modica2018}, authors apply Memguard at the hypervisor level, and Crespo et al. \cite{Crespo2018} follow a similar line by applying control theory to implement a feedback control scheme. Others have analyzed the efficacy of QoS regulators in minimizing IO-originated DRAM contention \cite{serranocases2021, Zini2022}.

\mypara{Co-scheduling}. The main insight of this approach is to co-schedule workloads in such a way that minimizes interference. RT-Gang \cite{Ali2019} proposes a novel gang-scheduling policy combined with memory bandwidth regulation. The PRedictable Execution Model (PREM) \cite{Pellizzoni2011} divides tasks into memory access and compute phases and co-schedules them accordingly so that the former do not overlap for critical workloads. Kloda et al. \cite{kloda2019deterministic} apply this idea at the hypervisor level.

\section{Conclusion}

In this paper, we presented the design, implementation, and evaluation of IRQ coloring. This mechanism aims to minimize interrupt-generated interference and provide intermediate guarantees for medium-criticality workloads. The core concept consists of deactivating "colored" interrupts if the \ac{QoS} of critical workloads drops below a specific threshold. The prototype was evaluated on a high-performance multicore platform (Xilinx ZCU102). Results demonstrated negligible performance overhead, i.e., $<$ 1\% for a sampling period of 100 microseconds, and reasonable throughput guarantees for medium-critical workloads. We believe that IRQ coloring is orthogonal to other state-of-art techniques, presents predictability and intermediate guarantees advantages, and can be implemented in the new generation of real-time Arm processors (e.g., Cortex-R52).

%\newpage

\section*{Acknowledgment}

This work has been supported by Huawei Pisa Research Center and by FCT - Fundação para a Ciência e Tecnologia grants 2022.13378.BD, 2020.04585.BD, SFRH/BD/138660/2018. This work has also been partially supported by FCT within the R\&D Units Project Scope UIDB/00319/2020.

\bibliographystyle{IEEEtran}
\bibliography{irq_coloring.bib}

\vfill

\end{document}